\documentclass[3p]{elsarticle}

\usepackage{amsmath,amsthm,amsfonts,amssymb,bm}
\usepackage[linktocpage=true]{hyperref}
\usepackage[final]{showkeys}
\usepackage[utf8]{inputenc}
\usepackage[english]{babel}
\usepackage{color}

\def\p{{\partial}}
\def\Re{\mathop{\mathcal R e}\limits}
\def\Im{\mathop{\mathcal I m}\limits}

\allowdisplaybreaks

\begin{document}

\begin{frontmatter}

\title{Theory of stochastic Laplacian growth}
\author{Oleg Alekseev}
\author{Mark Mineev-Weinstein}

\address{%
International Institute of Physics,\\ Federal University of Rio Grande do Norte, 59078-970, Natal, Brazil
}%

\begin{abstract}
	We generalize the diffusion-limited aggregation by issuing many randomly-walking particles, which stick to a cluster at the discrete time unit providing its growth. Using simple combinatorial arguments we determine probabilities of different growth scenarios and prove that the most probable evolution is governed by the deterministic Laplacian growth equation. A potential-theoretical analysis of the growth probabilities reveals connections with the tau-function of the integrable dispersionless limit of the two-dimensional Toda hierarchy, normal matrix ensembles, and the two-dimensional Dyson gas confined in a non-uniform magnetic field. We introduce the time-dependent Hamiltonian, which generates transitions between different classes of equivalence of closed curves, and prove the Hamiltonian structure of the interface dynamics. Finally, we propose a relation between probabilities of growth scenarios and the semi-classical limit of certain correlation functions of ``light'' exponential operators in the Liouville conformal field theory on a pseudosphere.

\end{abstract}

\end{frontmatter}

	\section{Introduction}

The problem of pattern formation in non-equilibrium growth processes has attracted great interest since the 1980s. The growth occurring in systems where diffusion plays an important role deserves particular attention. It describes such processes as formation of river networks, frost on glasses, growth of bacterial colonies, electric breakdown, dendritic crystals in rocks, polymerization, to name just a few. Patterns occurring in these systems are in general scale-invariant multi-branched fractal clusters. Remarkably, the complex shapes, are related to the instability of the diffusion-limited process, are universal, i.e.\ their long-range properties depend very slightly on the details of the interaction between particles. They have some general features providing a possibility to get insight into complex pattern formation by means of simple aggregation models. The best known example is the diffusion-limited aggregation (DLA) proposed by T.~Witten and L.~Sander in 1981~\cite{WS}. Due to the universality this model is widely applicable to study formation of complex patterns and their fractal properties in non-equilibrium and highly unstable growth processes. However, despite its simplicity many aspects of DLA remain puzzling.

DLA describes aggregation of small particles undergoing Brownian motion to form clusters. The density of these particles is assumed to be low, so that the aggregation process occurs by attaching one particle to the growing cluster per time unit. This process can be described by the following model. Consider a seed particle at the origin of a two-dimensional lattice. Another particle is launched from a distant source and is allowed to walk randomly. Eventually the second particle will stick irreversibly to the seed thus occupying one of lattice sites adjacent to the origin. Then the third randomly walking particle is launched from the same source. It moves around the lattice until it sticks to the two-particle cluster, and so forth. This process leads to a highly-branched fractal cluster. Remarkably all clusters grown in this way are mono-fractals with the same Hausdorff (fractal) dimension, $d_F = 1.71 \pm 0.01$, obtained numerically, which appears to be robust and universal~\cite{Halsey}. This number is (almost) insensitive to the lattice structure, depends weakly on the geometry of the problem, and still is out of analytical reach.

Let us briefly recall a mathematical description of DLA. The growth of the cluster is specified by the set of growth probabilities, i.e.\ probabilities that some perimeter site is next to be added to the cluster. To be more specific, let $u(x,t)$ be the probability that the random walker reaches the point~$x$ at time~$t$. The field $u$ obeys the diffusion equation, $\partial u/ \partial t = \nabla^2 u$. Assuming that the growth is sufficiently slow, such that $\partial u/ \partial t$ can be neglected, the probability satisfies the Laplace equation, $\nabla^2 u=0$, with a condition $u=0$ at the cluster boundary. The growth velocity of the interface is proportional to the gradient of the probability density at the boundary, $\nabla_n u$, where $n$ is the unit normal outside the growing cluster. Remarkably, the same set of equations describes a completely different (at first sight) non-equilibrium process, namely, the Laplacian growth (LG) in a Hele-Shaw cell~\cite{LG}. In the Laplacian growth problem the pressure field satisfies the Laplace equation with a constant pressure at the boundary (if to neglect surface tension), while the velocity of the interface is proportional to the pressure gradient. It should be noted that DLA corresponds to \textit{unstable} Laplacian growth so that the inviscid fluid pushes the viscous one. The latter problem is ill-posed in a mathematical sense, because evolution of an arbitrary interface often leads to generation of cusps within a finite time~\cite{Cusps}. To handle this problem one can use a regularization, such as surface tension. However, a wide class of non-singular logarithmic exact solutions are known to exist even for zero surface tension~\cite{log-0,log-1}.

Another remarkable feature of the Laplacian growth is formation of universal asymptotic shapes observed in experiments. The best known example is the Saffman-Taylor finger propagating in a long rectangular Hele-Shaw cell~\cite{ST}. A continuous family of possible fingers labeled by $\lambda$ (the ratio of the finger width to the channel width) can be easily obtained~\cite{ST}. However, experimentally only the unique pattern with $\lambda=1/2$ is observed. This fact raised a non-trivial problem of selection of a single (observed) pattern from infinitely many solutions. Historically, the problem was addressed first by including surface tension and applying the WKB-like technique (``the asymptotic beyond all orders'') to  study the low-surface-tension limit~\cite{SurfT,Kruskal}. More recently it was shown that the selection problem can be  solved using exact solutions of Laplacian growth even without surface tension~\cite{Mark-Selection}, given the shape of the initial interface which evolves deterministically. 

However, the Laplacian growth is known to be a highly unstable, dissipative, non-equilibrium, and non-linear phenomenon. It is expected that the shape of the interface can change stochastically during the evolution. Therefore, it seems reasonable to consider stochastic Laplacian growth, and its deep connection with DLA provides a clue to the problem. The well-known attempt to address the problem from this point of view was made by M.~Hastings and L.~Levitov by means of iterating stochastic conformal mappings~\cite{Levitov}. In their approach the cluster is grown by adding a small semicircular piece to the interface with a probability, which depends on the local electric field (that is a gradient of pressure) in terms of the Laplacian growth. In the continuum limit, where the attached bumps are infinitesimally small, the growth is described by the deterministic Laplacian growth equation.

In this paper we develop another approach to the stochastic pattern formation in the Laplacian growth, which allows one to study the grown clusters \emph{analytically}. The model we propose is simple and serves as a bridge between the Laplacian growth and DLA. It is assumed that $K$ particles (instead of one) are deposited from a distant source per time unit. The particles are assumed to be uncorrelated and move around until they stick to the cluster interface. As soon as all of them become a part of the cluster other $K$ particles are released from a distant source, and the process continues in this way. 

For $K=1$ the model is equivalent to the ordinary DLA, while for large $K$ it describes the stochastic Laplacian growth. In the limit, when the particles' sizes go to zero, we recover the deterministic Laplacian growth dynamics, governed by a single hydrodynamical source at infinity. It is convenient to consider roughly square-shaped particles, whose sides are formed by segments of the equipotential and field lines. Being stuck to the interface each time step, they form an external layer of the emergent domain. Since their shapes are specified, the evolution of the cluster is completely determined by the distributions of the attached particles along the domain boundary. Therefore a statistical description of different growth processes becomes possible.

The paper is organized as follows. In Section~\ref{2} we fix the notation and review classical Laplacian growth and its integrable structure. In Section~\ref{3} we introduce the stochastic Laplacian growth model and define the probability of the clusters. Besides, we prove that the classical (deterministic) limit of the model describes the deterministic Laplacian growth governed by a single hydrodynamical source at infinity. In other words, we introduce the {\it variational approach} to the Laplacian growth and determine the {\it action}, which extremum gives the equation for dissipative motion of the boundary of the domain. So far the Laplacian growth equation was derived only as the approximation of viscous hydrodynamics. The rest of the section is dedicated to other (non-classical) growth scenarios and their statistical weights, which are related to the {\it entropy} (number of states with specific parameters) of growing clusters. Surprisingly, these purely probabilistic expressions can be identically transformed to the electrostatic energy of a uniformly distributed charge inside the stochastically grown layer per elementary time unit. Thus, the statistical weights of the clusters obey the usual Gibbs-Boltzmann statistics, which, as a rule, is not applied out from equilibrium. The relations between our results and some topics of the modern mathematical physics, such as the theory of random matrix and the K\"ahler geometry, are also briefly addressed. In Section~\ref{4} we reveal the Hamiltonian structure of the interface dynamics and interpret the Laplacian growth equation as the Hamilton's equation for a certain dynamical system. In Section~\ref{5} we indicate a possible connection between our results and the Liouville conformal field theory. Finally, we draw our conclusions and mention some open questions.

	\section{Classical Laplacian growth}\label{2}

\subsection{Standard formulation and conformal description of the Laplacian growth}

The Saffman-Taylor problem (known also as the Laplacian growth) describes motion of the interface between two incompressible fluids with different viscosities in a Hele-Shaw cell. Equations for interface dynamics take extremely simple form in the limit when one fluid is inviscid and surface tension is zero. We will refer to inviscid and viscous fluids as water and oil respectively. We also assume that the water droplet, $D^+$, is simply connected and contains the origin, while the oil domain will be $D^-\equiv D=\mathbb C\setminus D^+$. In the water (with zero viscosity) the pressure field is a constant. In the oil, in turn, the fluid velocity is governed by the Darcy's law (eq.~\eqref{Darcy-law} below). Let $z=x+i y$ be a complex coordinate on the plane. Let ${\bf v}(z,t)$ and $p(z,t)$ be the velocity and the pressure field in the oil. Then the Darcy's law expresses a simple proportional relationship between ${\bf v}(z,t)$ and the pressure gradient:
\begin{equation}\label{Darcy-law}
	{\bf v}(z,t)=-\text{grad}\, p(z,t),\qquad z\in D,
\end{equation}
where we set to~$1$ the proportionality positive constant, which depends on viscosity and the thickness of the Hele-Shaw cell. In the absence of surface tension the pressure at the interface vanishes,
\begin{equation}\label{p-boundary}
	p(z,t)\Big|_{\p D}=0.
\end{equation}
A continuity condition equates the normal boundary velocity to the normal fluid velocity, $V_n(\zeta, t)$, at the interface:
\begin{equation}\label{v-boundary}
	V_n(\zeta,t)=-\p_n p(\zeta,t),\qquad \zeta\in \p D,
\end{equation}
where $\p_n$ is a normal derivative.

Consider a set of sources with rates $q_1,q_2\dotsc,q_M$ at the points $A_1,A_2\dotsc,A_M\in D$. Near the sources the pressure diverges logarithmically. Since oil is incompressible, its velocity is divergence-free, $\text{div}\, {\bf v}=0$, everywhere in $D$ except the source positions. Thus, because of ~\eqref{Darcy-law}, pressure obeys the Laplace equation:
\begin{equation}\label{p-many}
	\Delta\, p(z,t)=\sum_{m=1}^M q_m\delta^{(2)}(z-A_m),\qquad z,A_m\in D,
\end{equation}
where $\Delta=4\p\bar{\p}$ is a Laplace operator on the plane, and $\delta^{(2)}(z)$ is a two-dimensional delta-function.

The equations~\eqref{p-boundary} and~\eqref{p-many} completely specifies pressure and therefore the boundary velocity in~\eqref{v-boundary}. The unique solution for $p$ can be written as the linear combination of the Green's functions of $D(t)$,
\begin{equation}\label{p-G}
	p(z,t)=\sum_{m=1}^M \frac{q_m}{2\pi} G_{D(t)}(z,A_m).
\end{equation}

The Green's function is symmetric in its arguments, vanishes at the boundary, $G(z,\zeta)=0,\ \zeta\in \p D$, and satisfies the Laplace equation: $\Delta G(z,z')=2\pi \delta^{(2)}(z-z')$, where we omitted the domain's label $D$ for brevity. As a function of $z$ it is harmonic everywhere in $D$ except the point, $z=z'$, where it diverges logarithmically: $G(z,z')=\log|z-z'|+reg$. Also, $G(z,\infty)=-\log|z|+reg$, as $z\to\infty$. The Green's function has a simple electrostatic interpretation.  Suppose the boundary of the domain, $\p D$, is a grounded conducting wire~\footnote{The potential equals zero on it} and at some point, $z\in D$, there is a unit charge, exerting an electric field of force derived from a logarithmic potential. It is convenient to decompose the Green's function into two parts:
\begin{equation}\label{G+def}
	G(z,z')=\log|z-z'|+G^-(z,z').
\end{equation}
The first function at the right hand side, $G^+(z,z')=\log|z-z'|$, is a potential at ~$z'$ created by the point-like charge located at~$z$. The second function, $G^-(z,z')$, is a potential of countercharges induced at the equipotential domain's boundary. The Green's function can be written explicitly using the Riemann mapping theorem. According to the theorem, the external domain $D$ is conformally equivalent to the compliment $B=\mathbb C\setminus B_+$ of the unit disk $B_+=\{w\in \mathbb C:|w|<1\}$. Let the complex analytic function,
\begin{equation}\label{z-map}
	z=f(w):\quad \mathbb C\setminus B\to D,
\end{equation}
realizes the correspondence. The map is fixed by the conditions: $f(\infty)=\infty$ and $f'(\infty)>0$. Then, the Green's function in $D$ is
\begin{equation}\label{G-def}
	G(z,z')=\log\left|\frac{w(z)-w(z')}{1-w(z)\overline{ w( z')}}\right|,
\end{equation}
where $w=w(z)$ is the inverse function to $z(w)$ and bar stands for complex conjugation.

The equation of motion of the interface follows from~\eqref{v-boundary} via the conformal map $z=f(w)$, as the pressure is given by~\eqref{p-G}. The normal velocity in terms of the conformal map equals $V_n=\Im(\bar V \tau)=\Im(\bar z_t z_l)$, where $\tau=dz/|dz|$ is a unit tangent vector, and $l=\int|dz|$ is an arc-length of the boundary. After changing a parametrization of the boundary from the arc-length to the angle at the $w$-plane, $V_n$ takes the form:
\begin{equation}\label{vcl-def}
	V_n(e^{i\phi},t)=|f'(e^{i\phi})|^{-1}\Im\left(\p_t\overline{f(e^{i\phi})}\p_\phi f(e^{i\phi})\right),
\end{equation}
where $e^{i\phi}=w(\zeta)$ is a pre-image of the boundary. From~\eqref{v-boundary} and~\eqref{G-def} we have
\begin{equation}\label{pnG-n}
	\p_n G(\zeta,A_m)=|f'(e^{i\phi})|^{-1}\Re\frac{e^{i\phi}+a_m}{e^{i\phi}-a_m},
\end{equation}
where $a$ is the inverse pre-images of $A=f(1/\bar a)$. From~\eqref{v-boundary}, \eqref{p-G}, \eqref{vcl-def}, and~\eqref{pnG-n} we obtain
\begin{equation}\label{LG-eq}
	\Im\left(\p_t \overline{f(\phi,t)}\p_\phi f(\phi,t)\right)=\sum_{m=1}^M\frac{q_m}{2\pi}\Re\frac{e^{i\phi}+a_m}{e^{i\phi}-a_m},
\end{equation}
which is a famous Laplacian growth equation for interface evolution in the presence of several sources~\cite{Kufarev}. For a single source at infinity, it takes a familiar form~\cite{Galin,Polubarinova}:
\begin{equation}\label{LG-1}
	\Im\left(\p_t \overline{f(\phi,t)}\p_\phi f(\phi,t)\right)=\frac{q}{2\pi}.
\end{equation}

Remarkably, the nonlinear equations~\eqref{LG-eq} and \eqref{LG-1} have wide classes of exact solutions. The simplest examples are  polynomial and rational solutions which, however, exhibit cusp-like singularities at the interface in a finite time~\cite{Cusps}. This signifies importance of surface tension effects at highly curved parts of the interface. Singularity free interface dynamics with zero surface tension is described by logarithmic solutions~\cite{log-0,log-1} (and even more general multi-cut ones~\cite{Abanov}).

Another remarkable property of the Laplacian growth is an existence of an infinite set of conservation laws discovered by S.~Richardson~\cite{Ri72}. Introducing the set of internal and external harmonic moments,
\begin{equation}\label{tk-vk-def}
	\begin{gathered}
	v_k=\frac{1}{\pi}\int_{D^+} z^k d^2z,\quad v_0=\frac{2}{\pi}\int_{D^+}\log|z|d^2z,\\
	t_k=-\frac{1}{\pi k}\int_{D} z^{-k}d^2z, \quad t_0=\frac{1}{\pi}\int_{D^+} d^2z,
	\end{gathered}
\end{equation}
the Richardson's theorem relates the time evolution of $t_k$ with the locations and rates of the sources in $D$:
\begin{equation}\label{dt-def}
	\frac{d t_k}{d t}=\frac{1}{\pi}\sum_{m=1}^M q_m \frac{A_m^{-k}}{k},\qquad \frac{d t_0}{d t}=\frac{1}{\pi}\sum_{m=1}^M q_m,
\end{equation}
In particular, for the LG with a single source at infinity~\eqref{LG-1} all $t_k$ $(k>0)$ are constants, while the domain's area, $\pi t_0$, growth linearly with time. If the sources rates in~\eqref{dt-def} are constants, the time derivatives of $t_k$ are conserved. Thus, the Laplacian growth possesses an infinite number of integrals of motion, which is a distinct feature of integrability. Since equations~\eqref{dt-def} can be integrated, the Laplaican growth problem reduces to the inverse potential problem of  reconstructing the domain by its Newtonian potential~\cite{PS},
\begin{equation}\label{Phi-def}
	\Phi(z)=-\frac{2}{\pi}\int_{D^+} d^2z'\log|z-z'|,
\end{equation}
which admits the following Taylor series expansions for $z\to \infty$ and $z\to0$ respectively:
\begin{equation}\label{Phi-minus-def}
	\Phi^-(z)=-2t_0\log|z|+2\Re\sum_{k>0}\frac{v_k}{k}z^{-k},\qquad
	\Phi^+(z)=-|z|^2-v_0+2\Re\sum_{k>0} t_kz^k.
\end{equation}
Here $v_k$ and $t_k$ are the harmonic moments introduced in~\eqref{tk-vk-def}. The continuity of the potential and its gradient at the interface is expressed by the conditions: $\Phi^+(z)=\Phi^-(z)$, and $\p_z\Phi^+(z)=\p_z\Phi^-(z)$ for $z\in \p D$. This equations allow to reconstruct $\Phi^+(z)$ from $\Phi^-(z)$ (or vice versa) and, therefore, to treat the internal harmonic moments, $v_k$ ($k>0$) and $t_0$ as functions of external ones, $t_k$ ($k>0$), and $v_0$. Usually, it is more convenient to chose $t_0$ instead of $v_0$ as an independent parameter. Therefore, any functional on the domain, e.g\ the potential~\eqref{Phi-def}, can be considered as a function of the external harmonic moments, $t_k$ and $\bar t_k$, and the area $\pi t_0$. Under certain conditions  they completely determine the domain.

	\subsection{Schwarz function description of the Laplacian growth}

The LG equation~\eqref{LG-eq} can be integrated in terms of the {\it Schwarz function}~\cite{Schwarz}, which turns out to be of great use in different contexts. To introduce the Schwarz function of a certain contour $\Gamma$, defined by the equation $F(x,y)=0$, we substitute $x=(z+\bar z)/2$ and $y=(z-\bar z)/2$, and solve the resulting equation locally for $\bar z$ in terms of $z$:
\begin{equation}\label{Sf-def}
	\bar z=S(z).
\end{equation}
Then, the Schwarz function is defined as the analytic continuation of $S(z)$ away from the curve. The conformal map and the Schwarz function are connected in the following way: $z=f(w)$ and $S=\bar f(1/w)$.

The {\it Herglotz's theorem}~\cite{Herglotz} establishes a correspondence between singularities of the Schwarz function of a curve, $\Gamma$, and of the conformal map from the unit circle to $\Gamma$. Let $a$ be a singularity of $f(w)$ inside the unit circle, so $f(w) = A/(w-a)^m$ near $a$, and $m>0$ for integer $m$. Then the Schwarz function has a singularity of the same kind at $b\in D$ with a coefficient $B$, which is related to $(a,A)$~as: $b=f(1/\bar a)$ and  $\bar B=A(-a^2f'(1/\bar a))^m$, where $m$ is a multiplicity of the pole, rational number or zero if $a$ is a pole, algebraic branch point or logarithmic singularity respectively. 

For analytic contours the Schwarz function is well defined in the strip-like neighborhood of the curve and can be decomposed into the sum $S(z)=S^+(z)+S^-(z)$, where the functions $S^\pm(z)$ are regular in $D^\pm$. Their expansions near the origin and infinity have a form:
\begin{equation}\label{Spm-def}
	S^+(z)=\sum_{k=1}^\infty kt_k z^{k-1},\quad S^-(z)=\frac{t_0}{z}+\sum_{k=1}^\infty v_k z^{-k-1},
\end{equation}
where the coefficients are the harmonic moments~\eqref{tk-vk-def}.
If $t_k$ change according to Richarson's theorem~\eqref{dt-def} the time evolution of $S^+(z)$ is
\begin{equation}\label{dS+}
	\frac{d}{d t} S^+(z,t)=-\frac{1}{\pi}\sum_{m=1}^M\frac{q_m}{z-A_m}.
\end{equation}
Thus, singularities of $S^+(z,t)$ are in one-to-one correspondence with the {\it operating sources} $q_m$.  The initial $S^+(z,0)$ can  also have singularities, which are the {\it frozen sources} operating in the past,  $t<0$. Typical singularities of the Schwarz function in the water domain, $S^-(z)$, are the branch points of order two with cuts between them.

The Darcy's law~\eqref{Darcy-law} relates the time evolution of the Schwarz function to pressure in $D$. From $\bar z=S(z,t)$, by the chain rule we obtain:
\begin{equation}\label{bar-v}
	\overline V=\dot S+S' V,
\end{equation}
where $V=\dot z$. Introducing the unit tangent and normal vectors as $\tau=dz/|dz|$ and $n=-i\tau=1/i\sqrt{S'}$ respectively, where $|dz|=\sqrt{S'}dz$ is an arc-length, absence of the tangential velocity of the interface takes the form: $V_\tau=\Im(\overline V n)=0$. Since $S'=1/\bar S'$ we transform the latter condition to $\bar V+S' V=0$. Together with~\eqref{bar-v} it results in the relation: $\overline V=\dot S/2$. The Darcy's law can be also written as $V=-2\bar \p p$, or, equivalently, $\overline{V}=-2\p p$. Thus, we obtain:
\begin{equation}\label{SW-eq}
	\dot S(z,t)=2 W'(z,t),
\end{equation}
where dot and prime are time and space partial derivatives respectively, and $W(z,t)$ is the \emph{complex potential}, such that $p=-\Re W$. Eq.~\eqref{SW-eq} is equivalent to the Laplacian growth equation. Indeed, the normal boundary velocity, $V_n=\Re(\overline V n)$, can be written in terms of the Schwarz function:
\begin{equation}\label{vn-def-S}
	V_n(\zeta,t)=\frac{\dot S(\zeta,t)}{2i \sqrt{S'(\zeta,t)}},\quad \zeta\in \p D.
\end{equation}
Being projected onto the unit normal vector, the LG equation~\eqref{SW-eq}, together with~\eqref{vn-def-S}, is equivalent to~\eqref{v-boundary}:
\begin{equation}\label{vn-sum-G}
	V_n(\zeta,t)=-\sum_{m=1}^M \frac{q_m}{2\pi}\p_n G(\zeta,A_m),\quad \zeta\in \p D.
\end{equation}

The primitive of the Schwarz function (the so-called \textit{generating function}), $\Omega(z)=\int^z S(z')$, will play an important role in what follows. Let us briefly recall its main properties. From the series expansion~\eqref{Spm-def} we obtain: $\Omega(z)=\Omega^+(z)+ \Omega^-(z)- \frac12 v_0$, where $\Omega^\pm(z)$ are analytic in $D^\pm$ respectively:
\begin{equation}\label{Omega-series}
	\Omega^+(z)=\sum_{k>0}t_kz^k,\qquad
	\Omega^-(z)=t_0\log z-\sum_{k>0}\frac{v_k}{k}z^{-k}.
\end{equation}
Comparing~\eqref{Omega-series} with~\eqref{Phi-minus-def}, the generating function is related with the electrostatic potential as follows:
\begin{equation}\label{phi-pm}
	\Phi^-(z)=-2\Re \Omega^-(z),\qquad 
	\Phi^+(z)=2\Re \Omega^+(z)-v_0-|z|^2.
\end{equation}
Because of the logarithm in the Taylor series expansion of $\Omega^-(z) $, the generating function is multi-valued while its real part if well defined. In particular, $\Re \Omega(\zeta)=|\zeta|^2/2,\ \zeta\in \Gamma$. The generating function is harmonic in the internal domain with a logarithmic singularity at the origin.

	\subsection{Tau-function and integrability}

The harmonic moments, $t_k$ and $v_k$, are known to satisfy the  symmetry relations~\cite{Tau-int}:
\begin{equation}
	\frac{\p v_k}{\p t_n}=\frac{\p v_n}{\p t_k},\qquad \frac{\p v_k}{\p\bar t_n}=\frac{\p\bar v_n}{\p t_k}.
\end{equation}
These relations imply an existence of a real single-valued potential function, such that $v_k=\p\log \tau/\p t_k$ and $t_k=\p\log \tau/\p v_k$, where $\log \tau$ is the logarithm of the tau-function~\cite{Tau-calc},
\begin{equation}\label{log-tau-def}
	\log \tau=-\frac{1}{\pi^2}\int_{D^+}\int_{D^+}\log\left|\frac{1}{z}-\frac{1}{z'}\right|d^2zd^2z'.
\end{equation}
It can be represented as an infinite series in the harmonic moments of the domain. Taking into account the Taylor series expansion of the potential in the internal domain, $\Phi^+(z)$, and performing the term-wise integration, we obtain:
\begin{equation}\label{tau-series}
	\log \tau=\frac12 t_0v_0+\Re\sum_{k>0}t_k v_k-\frac{1}{2\pi}\int_{D^+} |z|^2 d^2z.
\end{equation}
The integral $\pi^{-1}\int_{D^+} |z|^2 d^2z=(1/2)t_0^2+\Re\sum_{k>0}k\, t_kv_k$ can also be expressed as a series in the harmonic moments, as follows from the Stokes formula. However, it is convenient to treat the last term separately.

Remarkably, the inverse conformal map, $w(z)$, can be written in terms of the tau-function:
\begin{equation}\label{w-z-def}
	\log w(z)=\log z-\p_{t_0}\left(\frac12\p_{t_0}+ \sum_{k>0}\frac{z^{-k}}{k}\p_{t_k}\right)\log \tau.
\end{equation}
Using~\eqref{G-def} we also conclude that
\begin{equation}\label{G-tau}
	G(z,z')=\log\left|\frac{1}{z}-\frac{1}{z'}\right|+\frac12 \nabla(z)\nabla(z')\log \tau,
\end{equation}
where the differential operator,
\begin{equation}\label{nabla-def}
	\nabla(z)=\p_{t_0}+\sum_{k>0}\left(\frac{z^{-k}}{k}\p_{t_k}+\frac{\bar z^{-k}}{k}\p_{\bar t_k}\right),
\end{equation}
is the variational derivative, acting in the space of functionals, $X=X(t_0,t_1,\bar t_1, \dotsc)$, on the domain, and related to the time derivative. By using the chain rule and the Richardson theorem~\eqref{dt-def}, we obtain:
\begin{equation}\label{dt-nabla}
	\frac{d}{d t}\cdot X=\sum_{m=1}^M \frac{q_m}{\pi}\nabla(A_m)\cdot X.
\end{equation}

The Laplacian growth equation~\eqref{SW-eq} follows from the definition of the tau-function, provided that the time evolution of the external harmonic moments is governed by Richardson's theorem~\eqref{dt-def}. Let us introduce the auxiliary potential,
\begin{equation}\label{tilde-Phi}
\tilde \Phi(z)=-(2/\pi)\int_{D^+}\log\left|\frac{1}{z}-\frac{1}{z'}\right|d^2z'=\nabla(z)\log \tau,
\end{equation}
where $z\in D$. Applying $\nabla(z')$ to both sides of~\eqref{tilde-Phi} and using~\eqref{G-tau}, we obtain:
\begin{equation}\label{nabla-Phi}
	\nabla(z')\Phi^-(z)=2G^-(z,z')+2\log|w(z')|,
\end{equation}
where we took into account that $\nabla(z)v_0=2\log|z|-2\log|w(z)|$, used the definition $G^-(z,z')=G(z,z')-\log|z-z'|$, and expressed the auxiliary potential, $\tilde \Phi(z)=\Phi(z)+ v_0+2t_0\log|z|$, in terms of the ordinary potential. Therefore from~\eqref{dt-nabla} and~\eqref{nabla-Phi} it follows:
\begin{equation}\label{Phi-G}
	\frac{d }{d t}\Phi^-(z)=2\sum_{m=1}^M\frac{q_m}{\pi}\left( G^-(z,A_m)+\log|w(A_m)|\right).
\end{equation}
Finally, differentiating both sides of~\eqref{Phi-G} w.r.t.\ $z$, we obtain the Laplacian growth equation~\eqref{SW-eq} for the analytic in $D$ part of the Schwarz function.

	\section{Stochastic Laplacian growth}\label{3}

\subsection{Random growth and classical trajectory}

The stochastic Laplacian growth describes aggregation of small particles undergoing Brownian motion to form clusters~\cite{Gruzberg}. In contrast to DLA, an arbitrary number $K$ of uncorrelated particles are simultaneously issued from a distant source per unit time $\delta t$. Each particle has a finite area $\hbar$, and its shape is a curvilinear quadrangle, $\sqrt\hbar\times\sqrt\hbar$, whose sides are formed by segments of the equipotential and field lines. The initial domain is a unit circle at the origin, so that $\hbar\ll1$. Undergoing Brownian motion the particles issued from the source reach the growing domain, $D(t)$, and stick to its boundary, $\p D(t)$, thus forming an external layer, $l(t)=D(t+\delta t)\setminus D(t)$, (with the area~$K\hbar$) of the advanced cluster, $D(t+\delta t)$. In the \emph{hydrodynamical limit} ($\hbar\to0$) we can introduce the rate of the source, $q$, as
\begin{equation}\label{qh-def}
	q \delta t= K\hbar.
\end{equation}

As soon as all particles became a part of the cluster, another portion of them is released from the source, and the process continues like this.
The issued particles attach to the interface $\Gamma$ with probabilities determined by the {\it harmonic measure}, $\mu(\zeta,z)$,  of the boundary, $\p D(t)$,~\cite{HMeasure}. It is defined as the probability for a Brownian particle issued at $z$ to hit the boundary at the given segment $|d\zeta|\in \Gamma$,
\begin{equation}\label{hm-def}
	\mu(\zeta,z)=-\frac{1}{2\pi}\p_n G(\zeta,z)|d\zeta|,
\end{equation}
where $G(z,z')$ is the (time-dependent) Green's function of $D(t)$~\eqref{G-def}. In electrostatics $\mu(\zeta,z)$ is a charge distribution induced at $\zeta\in \Gamma$ by a unit charge at $z$ to keep the interface equipotential. The study of the harmonic measure is facilitated by conformal mapping since $\mu(\zeta,z)$ is conformally invariant. Using the Green's function of the complement to the unit disk~\eqref{G-def},
the harmonic measure in the $w$-plane reads:
\begin{equation}\label{hm-w}
	\mu(e^{i\phi},w)=\frac{1}{2\pi}\Re\left(\frac{e^{i\phi}+w}{e^{i\phi}-w}\right)d\phi,
\end{equation}
where $d\phi=|d\zeta|/|f'(e^{i\phi})|$ is a little arc at the unit circle.

Let us divide the interface into $N\gg1$ segments with the arc-lengths $|d\zeta|=\sqrt\hbar$, such that the particles attach to the bins at the boundary.
If the number of particles is much larger than the number of bins, $K\gg N$, many of them, say $k(\zeta_n)$, can attach to the $n$-th bin, thus
forming a column with the height $\sqrt\hbar\cdot k(\zeta_n)$.
Since the particles stick to the boundary {\it probabilistically}, we can only consider the {\it probability} of a particular distribution of attached particles (the {\it layer}), $\textbf{k}=\{k(\zeta_1),\dotsc k(\zeta_N)\}$. This probability is given by the multinomial formula~\footnote{It is supposed that the distant source is located at infinity.}:
\begin{equation}\label{P-single}
	P(\mathbf{k})=K!\prod_{n=1}^N\frac{[\mu(\zeta_n,\infty)]^{k(\zeta_n)}}{k(\zeta_n)!}.
\end{equation}

If the growth continues until time $T\gg \delta t$ ($T/\delta t$ is integer and $i$ labels the time steps), the probability of a particular growth scenario is given by the product of the {\it conditional probabilities}, $P({\bf k}_{i+1})$, for the layer ${\bf k}_{i+1}$ to grow over the domain $D_i\equiv D(i \delta t)$:
\begin{equation}\label{P-step-def}
	\mathcal P(\Bbbk)=\prod_{i=1}^{T/\delta t}P({\bf k}_i),
\end{equation}
where $\Bbbk=\{\mathbf{k}_i\}_{i=1}^{T/\delta t}$ labels the \textit{growth scenarios}.

The ratio, $K/N$, is an important parameter of the model. It is the average number of particles attached to a single segment of the boundary per time unit. When $K/N$ is small, we drop a few particles onto the cluster per second. The DLA, when $K = 1$, can be called a quantum limit of a stochastic Laplacian growth, as correlations between particles in this case are maximal. The next particle always ``feels'' a slight change of the interface, caused by the previously landed particle, while both would be totally uncorrelated if emitted simultaneously. The opposite limit, $K/ N\gg1$, will be called ``classical'', as the interface moves deterministically.

In the large $K$ limit one can use the Stirling approximation to recast the probability~\eqref{P-single} in the form:
\begin{equation}\label{P-d-def}
	P(\mathbf{k})=\exp\left\{-\sum_{n=1}^N k(\zeta_n)\log \frac{k(\zeta_n)}{K \mu(\zeta_n,\infty)}\right\},
\end{equation}
where the terms $O\bigl(\log k(\zeta_n,i)\bigr)$ were omitted in the exponent. The exponent in~\eqref{P-d-def} is the {\it Kullback-Leibler entropy}~\cite{Kullback}, which measures a distance between two distributions, $k(\zeta_n)$ and $K \mu(\zeta_n,\infty)$. Since the total number of issued particles is $K$, there is a constraint:
\begin{equation}\label{k-constraint}
	\sum_{n=1}^N k(\zeta_n)=K.
\end{equation}

It is convenient to consider a continuum limit of the model, when $N\to\infty$ and $\delta t\to0$, and replace the sums over discrete labels, $n$ and $i$, by the integrals,
\begin{equation}\label{sum-int}
	\sum_{n=1}^N X(\zeta_n)=\oint_{\Gamma}\frac{|d\zeta|}{\hbar^{1/2}}X(\zeta),\quad \sum_{i=1}^{T/\delta t} X(i) =\int_0^T \frac{dt}{\delta t} X(t).
\end{equation}
In the continuum limit the probability~\eqref{P-d-def} takes the form:
\begin{equation}\label{P-climit}
	P(\mathbf{k})=\exp\left\{-\oint_{\Gamma} k(\zeta)\log\frac{k(\zeta)}{K\mu(\zeta,\infty)}\frac{|d\zeta|}{\sqrt\hbar}\right\}.
\end{equation}

The classical distribution ${\bf k}^{cl}$ of attached particles, which maximizes the probability, can be determined by varying~\eqref{P-climit} w.r.t.\ $k(\zeta)$ with the constraint~\eqref{k-constraint},
\begin{equation}\label{xcl-def}
	k^{cl}(\zeta)=K\mu(\zeta,\infty).
\end{equation}
Since the second variation of the functional at $k^{cl}(\zeta)$ is strictly negative, the classical trajectory provides the global maximum of the probability. It becomes exponentially sharp when $\hbar\to0$, so the fluctuations around the saddle point are suppressed. The probability of the classical trajectory equals 1, unless the next order terms of the Stirling approximation are taken into account,
\begin{equation}
	P(\mathbf{k}^{cl})=\exp\left\{ - \frac12\oint_{\Gamma}\log( K\sqrt\hbar\,|w'(\zeta)|)\frac{|d\zeta|}{\sqrt\hbar} \right\},
\end{equation}
where we expressed the harmonic measure in terms of the Green's function~\eqref{hm-def}, used $|d\zeta|=\sqrt\hbar$ and $\p_n G(\zeta,\infty)=-|w'(\zeta)|$ for $\zeta\in \Gamma$.

{\it The Laplacian growth equation} describes the evolution of the interface due to newly attaching particles at the boundary. Indeed, since the normal displacement of the boundary is $V_n(\zeta)\delta t=\sqrt \hbar\cdot k(\zeta)$, from~\eqref{xcl-def} we determine $V_n^{cl}(\zeta)$ corresponding to the classical trajectory:
\begin{equation}\label{vcl-inf}
	V^{cl}_n(\zeta)=-\frac{q}{2\pi}\p_n G(\zeta,\infty),
\end{equation}
where $q=K\hbar/\delta t$ is the rate of the source. As discussed, the Darcy's law~\eqref{vcl-inf} can be transformed in the Laplacian growth equation:
\begin{equation}\label{LG-eq-inf}
	\Im\left(\p_t\overline{f(\phi,t)}\p_\phi f(\phi,t)\right)=\frac{q}{2\pi}.
\end{equation}

{\it Thus, it turned out possible to derive the Laplacian growth equation directly from a variational calculus based on elementary combinatorics}. So far this equation was possible to deduce only from viscous hydrodynamics or kinetics~\cite{Pelce}. The Laplacian growth equation~\eqref{LG-eq-inf} describes the uniform growth of the initially unit disk. However, experiments in the Hele-Shaw cell are known to produce different complex universal patterns reflecting highly unstable and non-equilibrium nature of the Laplacian growth. We will address this problem in the next section.

	\subsection{Non-classical trajectories and virtual sources}

\begin{figure}[]
\centering
\includegraphics[width=0.6\columnwidth]{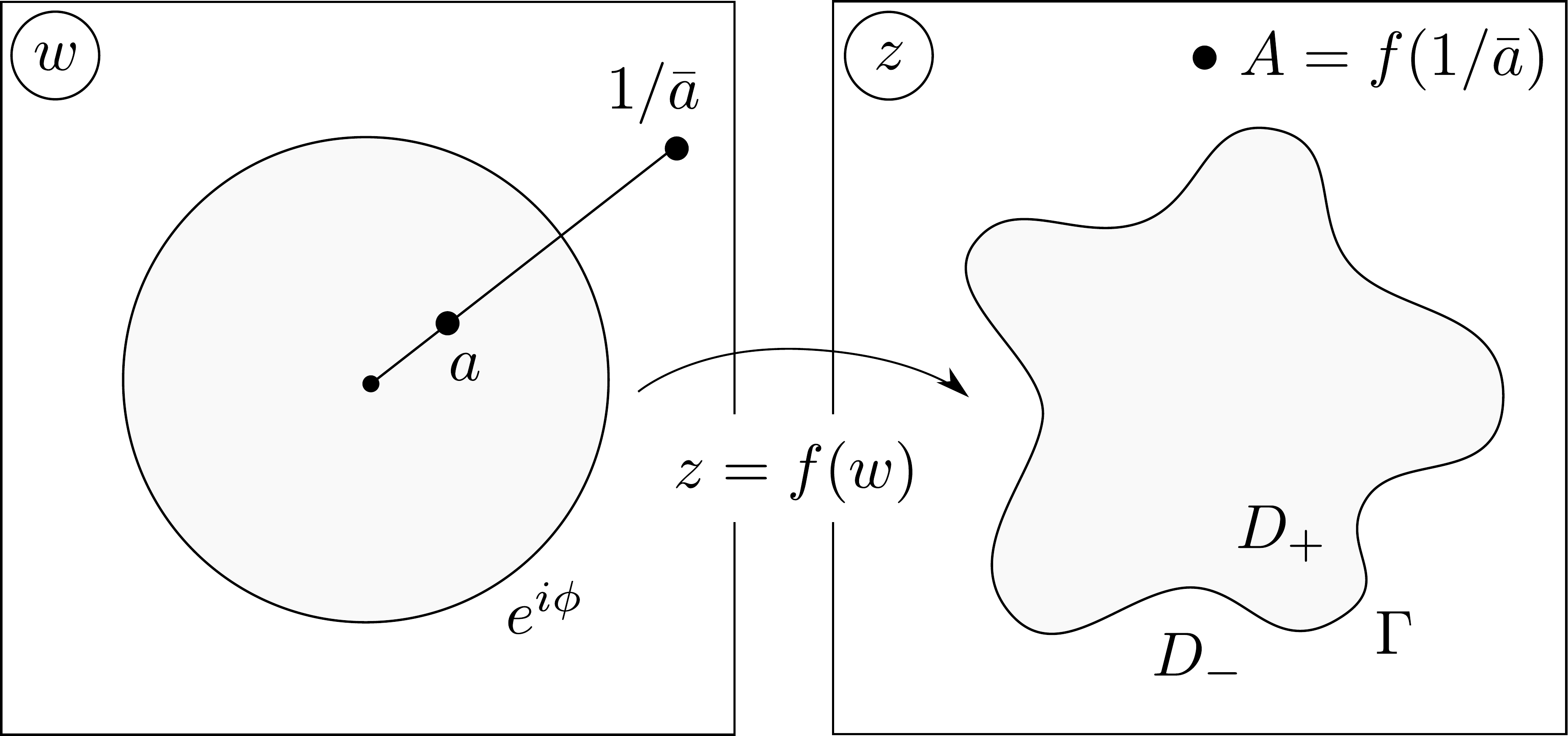}
    \caption{\label{map}Conformal map $z=f(w)$ from the exterior of the unit circle to $D_-$, so that $\infty = f(\infty)$, the conformal radius, $r = f'(\infty) > 0$, and $A = f(1/\bar a)$.}
\end{figure}

If the Brownian particles stick to the interface {\it stochastically}, the normal interface velocity deviates form its classical value~\eqref{vcl-inf}, resulting in the non-circular final domains. It is convenient to describe stochastic growth in terms of the {\it virtual sources}, which can be introduced as follows. Consider $M$ {\it independent} sources of Brownian particles $q_m$ located at the points $A_m\in D$ (instead of $\infty$), and repeat the probabilistic analysis of the previous section. Clearly, the probability distribution of attached particles, ${\bf k}_m$, issued from the $m$-th source is given by~\eqref{P-climit}, where the harmonic measure $\mu(\zeta,\infty)$ is replaced by $\mu(\zeta,A_m)$. Since the sources are {\it uncorrelated}, the overall probability distribution of attached particles is given by the product of $M$ multinomial probabilities,
\begin{equation}\label{tildeP}
	P(\mathbf{k})=\prod_{m=1}^M\exp\left\{-\oint_{\Gamma} k_m(\zeta) \log \frac{k_m(\zeta)}{K_m\mu(\zeta,A_m)}\frac{|d\zeta|}{\sqrt\hbar}\right\}.
\end{equation}
where $K_m=q_m \delta t/\hbar$ is the number of particles issued from the $m$-th source per time unit $\delta t$. Maximizing the probability~\eqref{tildeP} we determine the classical trajectory, $k^{cl}(\zeta)=\sum_m k_m^{cl}(\zeta)$, and the boundary velocity $V(\zeta)=\sum_m V_n^{(m)}(\zeta)$:
\begin{equation}\label{km-def}
	k_m^{cl}(\zeta)=K_m\mu(\zeta,A_m),\qquad V_n^{(m)}(\zeta)=-\frac{q_m}{2\pi}\p_n G(\zeta,A_m),
\end{equation}
The normal velocity obeys the Darcy's law, and therefore the interface dynamics follows the LG equation:
\begin{equation}\label{LGeq-M}
	\Im\left(\p_t \overline{f(\phi,t)}\p_\phi f(\phi,t)\right)=\sum_{m=1}^M\frac{q_m}{2\pi}\Re\frac{e^{i\phi}+a_m}{e^{i\phi}-a_m},
\end{equation}
where $a$ are the inverse pre-images of $A=f(1/\bar a)$ (see Fig.~\ref{map}). Since the same equation appears in the deterministic Laplacian growth problem~\eqref{LG-eq}, the sources $q_m$ of Brownian particles in the classical limit can be treated as the usual hydrodynamic sources.

\begin{figure}[]
\centering
\includegraphics[width=0.6\columnwidth]{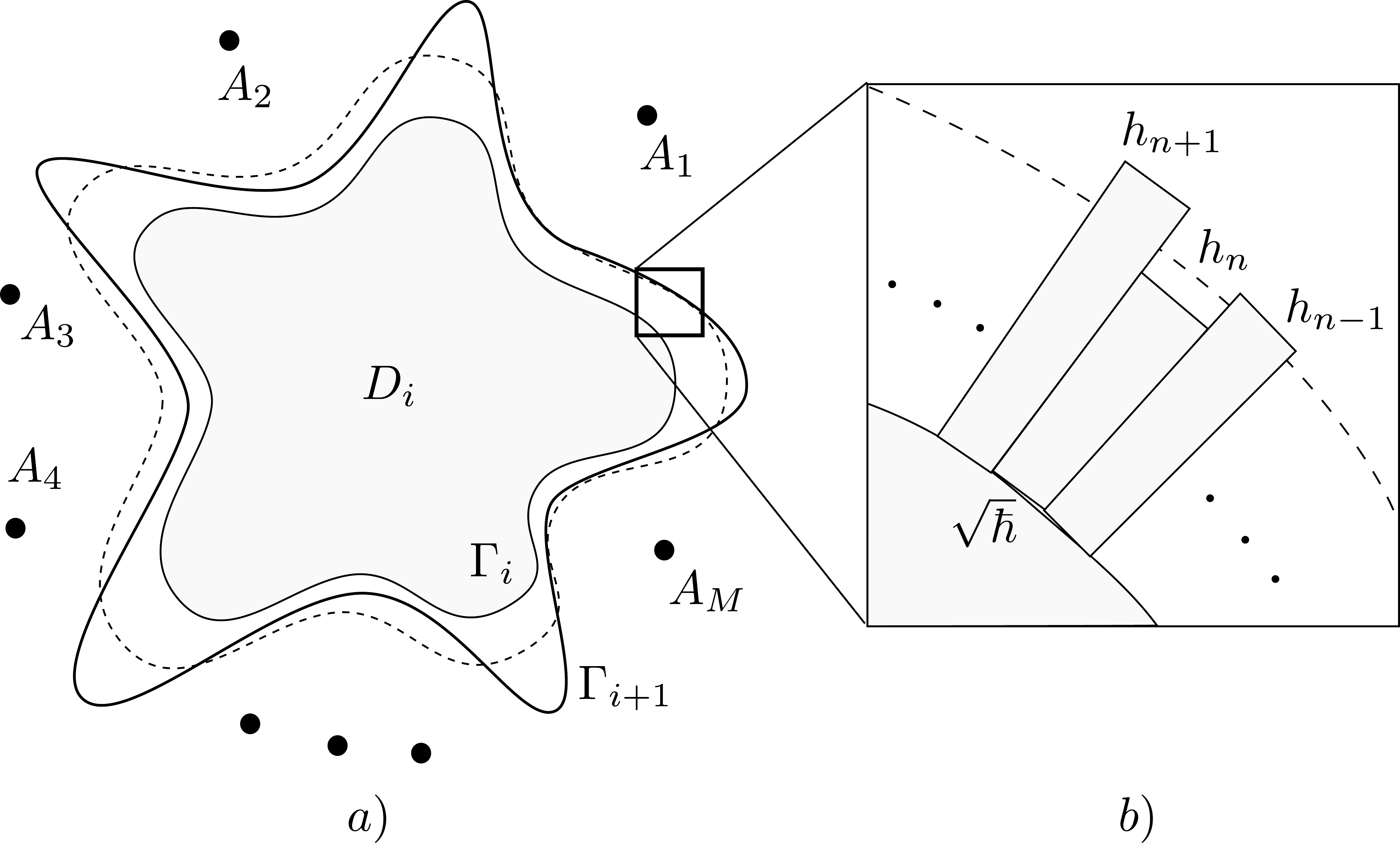}
    \caption{\label{layer}a) Stochastic growth of a single layer, $D_{i+1}/D_i$: Here a thin line is $\Gamma_i = \partial D_i$ formed during the first $i$ time units; a dashed line represents classical (deterministic) LG for a single source at $\infty$ during $(i+1)$-st unit, and a solid line, $\Gamma_{i+1}$ is an external boundary of a stochastic layer, $D_{i+1}/D_i$ grown per elementary time unit, $\delta t$. This stochastic layer is equivalent to a classical layer, grown in the presence of $M$ {\it virtual} sources located at $A_1$, $A_2$,\ldots, $A_M$. b) Three consecutive fragments of $\Gamma_i$, partitioned onto $N \gg1$ equal pieces of the size, $\sqrt \hbar$, after stochastic growth during the $(i+1)$-st time unit. The heights of grown columns equal to $h_{n } = \sqrt \hbar k_{n }$.}
\end{figure}

According to~\eqref{dS+}, any domain bears ``fingerprints'', $A_m\in D$ (singularities of the Schwarz function), left by the hydrodynamical sources operated earlier times. The two identities, $V_n(\zeta)=\dot S(\zeta)/2i\sqrt {S'(\zeta)}$ and $V_n(\zeta)\delta t=\sqrt \hbar\cdot k(\zeta)$, relate the variation of the Schwarz function of the boundary with the corresponding distribution of the attached particles ${\bf k}$. Thus, the ``non-classical'' deviations from the classical growth~\eqref{LG-eq-inf}, when the new poles (say $A_m$) on $S^+(z)$ emerge {\it stochastically}, can be attributed to the {\it virtual sources} at $A_m$, working in the classical regime. The distribution $k_m^{cl}(\zeta)$~\eqref{km-def} is non-classical w.r.t.\ the probability distribution attributed to the infinitely remote source~\eqref{P-climit} (see Fig.~\ref{layer}). Since the virtual sources are uncorrelated, the probability for occurring stochastic fluctuation of the interface (when the random poles $A_m$ of $S^+$ emerge) is given by
\begin{equation}\label{P-Dirichlet}
	P(\mathbf{k})=\mathcal N\exp\left \{-\sum_{m=1}^M K_m\oint_{\Gamma}\mu(\zeta,A_m)\log\left|\frac{\p_n G(\zeta,A_m)}{\p_n G(\zeta,\infty)}\right|\right\}.
\end{equation}
where we used~\eqref{km-def} to express $k^{cl}_m(\zeta)$ in terms of harmonic measure. By $\mathcal N$ we denoted the pre-factor, which emerges after taking $K_m/K$ (recall that $k_m^{cl}\sim K_m$) out of logarithm in~\eqref{P-climit},
\begin{equation}\label{N-def}
	\mathcal N=\exp\left\{K\log K-\sum_{m=1}^M K_m\log K_m\right\}.
\end{equation}
This factor has a clear statistical interpretation. It is a number of partitions of $K$ indistinguishable particles in the $M$ groups of $K_m$ particles, such that $K=\sum K_m$, i.e.\ $\mathcal N=K!/(K_1!K_2!\cdots K_{M}!)$, which becomes~\eqref{N-def} in the Stirling approximation.

\subsection{Growth probabilities and virtual sources}

The line integral along the boundary in the exponent of~\eqref{P-Dirichlet} represents the solution to the Dirichlet boundary problem. The Dirichlet boundary problem~\cite{Dirichlet} is to find a harmonic function in $D$, such that it is continuous up to the boundary, and equals a given function $u_0(\zeta)$ on the boundary. The solution to the problem is given by the Poisson integral formula~\eqref{D-formula}:
\begin{equation}\label{D-formula}
	u^H(z)=
	-\frac{1}{2\pi}\oint_{\Gamma} \p_n G(\zeta,z) u_0(\zeta)|d\zeta|\equiv
	\oint_{\Gamma} \mu(\zeta,z) u_0(\zeta),
\end{equation}

Let us first consider the numerator of the logarithm in~\eqref{P-Dirichlet}.  By virtue of the Laplacian growth equation for the Schwarz function~\eqref{SW-eq} together with~\eqref{dt-nabla} we obtain $|\p_n G(\zeta,A)|=|\nabla(A)S(\zeta)|$. Decomposing the Schwarz function, $S(\zeta)=S^+(\zeta) + S^-(\zeta)$,  using $\nabla(A) S^+(\zeta)=-1/(\zeta-A)$ in accordance with~\eqref{dS+}, and reducing the expression in the logarithm to a common denominator, the integral we consider can be recast in the form:
\begin{equation}\label{logG-2}
	\oint_\Gamma \mu(\zeta,A)\log|\p_n G(\zeta,A)|=\oint_\Gamma \mu(\zeta,A)\Bigl(\log\left|1-(\zeta-A)\nabla(A) S^-(\zeta)\right|-\log|\zeta-A|\Bigr).
\end{equation}
The harmonic extension of the logarithm is $\oint_\Gamma \mu(\zeta,z)\log|\zeta-A|=\log|z-A|-G(z,A)+G(z,\infty)$, where the Green's functions in the r.h.s.\ do not change the boundary value of $u_0(\zeta)=\log|\zeta-A|$, and make it harmonic in the whole domain, including the points $z=A$ and $z=\infty$. Therefore, the last term in the r.h.s.\ of~\eqref{logG-2} takes the form: $\oint_\Gamma \mu(\zeta,A)\log|\zeta-A|=-G^-(A,A)+G(A,\infty)$, where the function $G^-(z,z')=G(z,z')-\log|z-z'|$ is harmonic everywhere in $D$, except infinity.

The former logarithm in the r.h.s.\ of~\eqref{logG-2} can be transformed as follows. The Taylor series expansion of $S^-(z)$ in $D$~\eqref{Spm-def} is: $\nabla(A)S^-(z)=z^{-1}+\sum_{k>0}\nabla(A)v_k\, z^{-k-1}$. Therefore, $\log\left|1-(\zeta-A) \nabla(A) S^-(\zeta)\right|$ is a harmonic function of $\zeta$ everywhere in $D$, except infinity, where is diverges logarithmically, $-\log|\zeta|$. To make this function harmonic at infinity we add the Green's function, $-G(\zeta,\infty)$, which does not change the boundary's value. Note that at the point $z=A$, the cumbersome contribution of $(z-A)\nabla(A) S^-(z)$ under the logarithm vanishes and the {\it value} of the harmonically continued function at this point is represented by the Green's function, $-G(A,\infty)$, only. Thus, combining the contributions of both logarithms from~\eqref{logG-2}, we arrive to the following identity:
\begin{equation}\label{logG-3}
	\oint_\Gamma \mu(\zeta,A)\log|\p_n G(\zeta,A)|=G^-(A,A)-2G(A,\infty).
\end{equation}

Now, consider the denominator of the logarithm in~\eqref{P-Dirichlet}. Since $\p_n G(\zeta,\infty)=-|w'(\zeta)|$ at the boundary, and the expansion of $w(z)$ in $D$ is: $w(z)=z/r+\sum_{j\geq 0} p_j z^{-j}$, where $r$ is the conformal radius, the function $\log|w'(\zeta)|$ is harmonic everywhere in the external domain (including infinity) and continuous up to the boundary. Thus,
\begin{equation}\label{he-logw}
	\oint_{\Gamma}\mu(\zeta,A)\log|\p_nG(\zeta,\infty)|=\log|w'(A)|.
\end{equation}

Summarizing, the probability~\eqref{P-Dirichlet} of the layer ${\bf k}=\cup_{m=1}^M {\bf k}_m^{cl}$, which describes the stochastic dynamics~\eqref{km-def} or~\eqref{LGeq-M}, takes the form:
\begin{equation}\label{P-z-def}
	P(\mathbf{k})=\mathcal N\exp\left\{-\sum_{m=1}^M K_m\Bigl(G^-(A_m,A_m)-2G(A_m,\infty)-\log|w'(A_m)|\Bigr)\right\}.
\end{equation}
There are two natural ways to proceed: to rewrite $P(\mathbf{k}^{cl})$ in the $w$-plane, expecting to get some simplification, and/or to recast the probability in a certain functional on the domain in $z$-plane. Let us consider them in succession.

	\subsection{Growth probability in the $w$-plane}\label{S:P-m}

The transformation of the growth probability to the $w$-plane is facilitated by the conformal invariance of the harmonic measure, $ \p_n G(\zeta,A)=|w'(\zeta)|\p_n G(e^{i\phi},1/\bar a)$, where $a$ is the inverse pre-image $A=f(1/\bar a)$. Since $\p_n G(\zeta,\infty)=-|w'(\zeta)|$, the probability~\eqref{P-Dirichlet} can be recast in the form:
\begin{equation}\label{P-math}
	P(\mathbf{k})=\mathcal N\exp\left \{-\sum_{m=1}^M K_m\oint_{|w|=1}\mu(w,1/\bar a_m)\log\left|\p_n G(w,1/\bar a_m)\right|\right\}.
\end{equation}
Since $\p_n G(e^{i\phi},1/\bar a)=-(1-|a|^2)/|e^{i\phi}-a|^2$, the contour integral in~\eqref{P-math} is easily calculated to equal
\begin{equation}\label{mu-math-1}
	\oint_{|w|=1}\frac{1-|a|^2}{|w-a|^2}\log\frac{1-|a|^2}{|w-a|^2}\frac{dw}{2\pi i w}=-\log(1-|a|^2).
\end{equation}
This is a Robbin function~\cite{Gustafsson}, which has a clear elecrostatic interpretation. It is a potential at $a$ created by charges induced by a unit charge at $a$ on the unit circle, kept at zero potential. Since the points $A$ are singularities of the Schwarz function in $D$, the Herglotz theorem identifies the charges at $a$ with the singularities of the conformal map, $z=f(w)$, inside the unit disk. With help of~\eqref{mu-math-1} the probability~\eqref{P-math} can be transformed in the following neat expression:
\begin{equation}\label{Pw-math}
	P(\mathbf{k})=\mathcal N\exp \left\{\sum_{m=1}^M K_m\log\left(1-|a_m|^2\right)\right\}.
\end{equation}

{\it So, we conclude that the probability for occurring stochastic fluctuations during LG dynamics takes a remarkable form of the Gibbs-Boltzmann distribution}, upon transformation of the entropy~\eqref{tildeP} to electrostatic energy~\eqref{Pw-math}. This conclusion opens novel possibilities for analyzing non-equilibrium growth processes by tools of equilibrium statistical physics.

Another useful interpretation of the exponent of the probability can be obtained by recasting the single contour integral in~\eqref{P-math} in the double contour one:
\begin{equation}\label{Pmumu}
	P({\bf k})=\mathcal N \exp\left\{\sum_{m=1}^M K_m \oint_{|w|=1}\oint_{|w'|=1}\mu(w,1/\bar a_m)\log|w-w'|\mu(w',1/\bar a_m)\right\},
\end{equation}
Thus, the exponent in the probability is the electrostatic energy of self-interacting charge induced on the unit circle with density $\mu(w,1/\bar a)$, i.e.\ the non-equilibrium Dyson gas on a circle.

If the growth continues until time $T$, the probability of the growth process is the product of the conditional probabilities~\eqref{P-step-def}:
\begin{equation}\label{P-w-all}
	\mathcal P(\Bbbk)=\left(\prod_{i=1}^{T/\delta t}\mathcal N_i\right)\exp\left\{ \frac{1}{\hbar}\int_0^T \sum_{m=1}^M q_m(t)\log(1-|a_m(t)|^2)\,dt\right\},
\end{equation}
where the subscript $i$, as well as the argument $t$, indicate the time dependence of the corresponding variables. The rate $q(t)$ is a {\it characteristic function of the source}. When $q(t)=q \delta(t-t')$ the source operates during a single growth step only. For smooth characteristic functions the integrals of motion, $A_m=f(1/\bar a_m(t))$, provide time-dependence of $a_m(t)$. For example, the solution for the LG equation in presence of $M$ hydrodynamical sources $q_m$ located at the points $A_m$~\eqref{LG-eq}, is the rational function: $z=rw+\sum B_m/(w-a_m)$. Then, using the Herglotz theorem, and determining the area growth law, we obtain $2M+1$ equations for the unknowns $a_m(t)$, $B_m(t)$ and $r$:
\begin{equation}
	\begin{gathered}
	r/\bar a_m+\sum_{m=1}^M \frac{B_k \bar a_m}{1-a_k\bar a_m}=A_m,\qquad B_m=-Q_m a_m^2\left(r-\sum_{k=1}^M\frac{B_k \bar a_m^2}{(1-a_k\bar a_m)^2}\right),\\
	r^2-\sum_{m,k=1}^M \frac{\bar B_m B_k}{(1-\bar a_m a_k)^2}=1+\sum_{m=1}^M Q_m t,
	\end{gathered}
\end{equation}
where we took into account that the initial domain is a unit circle with the area $\pi$. By solving these equations, one can determine $a_m(t)$ and the associated probability of the growth process~\eqref{P-w-all}.
	\subsection{Growth probability in the $z$-plane}

In the $z$-plane one can transform the growth probability in the functional on the layer, which consists of all particles attached to the interface per time unit. Contribution from the numerator of the logarithm~\eqref{P-Dirichlet} can be transformed in the double contour integral along the interface. Using $\oint \mu(\zeta,A)\log|\p_n G(\zeta,A)|= -\oint\mu(\zeta,A)\log|\zeta-A|-G(A,\infty)$ and $\log|\zeta-A|=\oint \mu(\zeta',A)\log|\zeta-\zeta'|-G(A,\infty)$, we arrive to the identity:
\begin{equation}\label{A1-def}
	\oint_\Gamma \mu(\zeta,A)\log|\p_n G(\zeta,A)| 
	=-\oint_{\Gamma}\oint_{\Gamma}\mu(\zeta,A)\log|\zeta-\zeta'|\mu(\zeta',A).
\end{equation}

By virtue of the Stokes theorem we transform the double line integrals in a certain functional on the layer itself. However, an extra attention to logarithmic cuts inside the layer should be paid. By projecting both sides of LG equation~\eqref{SW-eq} to the unit normal, $n=1/\sqrt{-S'(\zeta)}$,  we obtain: $\p_n G(\zeta,A)=- n\cdot \nabla(A)S(\zeta)$, and $\nabla(z)$ is the differential operator~\eqref{nabla-def}.
Therefore, the double line integral in the r.h.s.\ of~\eqref{A1-def} reads:
\begin{equation}\label{logG-SS}
	\oint_{\Gamma} \mu(\zeta,A)\log|\p_n G(\zeta,A)|= -  \Re\oint_{\Gamma}\oint_{\Gamma}\frac{d\zeta}{2\pi i} \frac{d\zeta'}{2\pi i} \log(\zeta-\zeta')\nabla(A) S(\zeta)\nabla(A) S(\zeta') ,
\end{equation}
where we also used $n\cdot|d\zeta|=-id\zeta$.
Consider the  layer $l(t)=D(t)\setminus D(t-\delta t)$ of all particles issued from the source at $A$ and stuck to the boundary of $D(t-\delta t)$ during the growth step $\delta t$. Relating $\nabla(A)$ with the time derivative~\eqref{dt-nabla}, where $M=1$ is assumed, we obtain:
$\nabla(A)S_{t-\delta t}(\zeta)=(\pi/q\delta t)(S_t(\zeta)-S_{t-\delta t}(\zeta))$. Here $S_{t}(\zeta)$ is the Schwarz functions of the boundary curve $\p D(t)$. It is convenient to introduce the auxiliary integral,
\begin{equation}\label{I-def}
	I(z)=\frac{K \hbar}{\pi} \oint_{\Gamma(t-\delta t)} \mu(\zeta,A)\log|z-\zeta|,
\end{equation}
and recast it in the form:
\begin{equation}\label{I-def1}
	I(z)=\Re\oint_{\Gamma(t-\delta t)}\bigl(S_t(\zeta)-S_{t-\delta t}(\zeta)\bigr)\log(z-\zeta)\frac{d\zeta}{2\pi i}.
\end{equation}
Now, by moving the contour $\Gamma(t-\delta t)\to \Gamma(t)$ in the first integral in~\eqref{I-def1}, and taking into account the point $z$ inside the layer, we obtain:
\begin{equation}\label{SSint}
	\oint_{\Gamma(t-\delta t)}S_t(\zeta)\log|z-\zeta|\frac{d\zeta}{2\pi i}=-\Re\int_{[z,\zeta_0]}S_t(\zeta)d\zeta +
	\int_{\Gamma(t)\setminus\{\zeta_0\}}S_t(\zeta)\log|z-\zeta|\frac{d\zeta}{2\pi i}.
\end{equation}
Here $\zeta_0\in \Gamma(t)$  is the point, where the logarithmic cut $[z,\zeta_0]$ intersects the boundary, and the former integral in the r.h.s.\ of~\eqref{SSint} goes along both edges of the cut in a clockwise direction. Let us add and subtract the following difference of two integrals (along the lower, $[z_0,\zeta]$, and the upper, $[\zeta,z_0]$, edges of the cut) to $I(z)$:
\begin{equation}
	I_{cut}(z,\zeta_0)=\Re\left(\int_{[\zeta_0,z]}-\int_{[z,\zeta_0]}\right)S_{cut}(\zeta) \log(z-\zeta)\frac{d\zeta}{2\pi i}= -\Re\int_{[\zeta,z_0]}S_{cut}(\zeta) d\zeta,
\end{equation}
where $S_{cut}(\zeta)$ is the Schwarz function of the cut. One can apply the Stokes theorem to recast the sum of the integrals, $I_{\Gamma_{m}\setminus\{\zeta_0\}}+I_{cut}-I_{\Gamma_{m-1}}$, in the integral over the layer $l(t)$. Thus, we obtain:
\begin{equation}\label{I-def-2}
	I(z)=\frac{1}{\pi}\int_{l(t)}\log|z-z'|d^2z'
	-\Re\int_{[z,\zeta_0]}S_t(\zeta)d\zeta+\Re\int_{[z,\zeta_0]}S_{cut}(\zeta)d\zeta.
\end{equation}
Using the primitive of the Schwarz function, $\Omega (z)=\int^z S (z')dz'$, such that $\Re \Omega (z)=|z|^2/2$ for $z\in\Gamma$, we transform $I(z)$ in the form:
\begin{equation}\label{I-iint}
	I(z)=\frac{1}{\pi}\int_{l(t)}\log|z-z'|d^2z'-\frac{|z|^2}{2}+\Re \Omega_t(z).
\end{equation}
If $z\in \Gamma(t-\delta t)$ one can replace $|z|^2/2$ by $\Re \Omega_{t-\delta t}(z)$. Finally, by virtue of~\eqref{I-def-2} we recast the double contour integral~\eqref{logG-SS} in the double integral over the layer:
\begin{equation}\label{A1-int}
		-(K\hbar)^2\oint_{\Gamma(t-\delta t)} \mu(\zeta,A)\log|\p_n G(\zeta,A)|= 
		\int_{l(t)} \int_{l(t)} \log\left|z-z'\right|d^2zd^2z'- \pi \int_{l(t)}\mathcal A_{t}(z)d^2z,
\end{equation}
where $K\hbar$ is the area of the layer, and $\mathcal A(z)$ is the {\it modified Schwarz potential}~\cite{mSch}:
\begin{equation}\label{mSchwarz}
	\mathcal A(z)=\frac{|z|^2}{2}-\Re \Omega(z).
\end{equation}
Since $\Re \Omega(z)=|z|^2/2$ for $z\in \Gamma$, the modified Schwarz potential vanishes at the boundary. Besides, the first derivatives also vanish for all points of the boundary, $\p\mathcal A(z)=\bar\p \mathcal A(z)=0$ for $z\in \Gamma$. Thus, the contour $\bar z=S(z)$ is a saddle point of $\mathcal A(z)$. The derivation of~\eqref{A1-int} is straightforward, as $\int_{l(t)}\Re\Omega_{t-\delta t}=\int_{l(t)}\Re\Omega_t$.

The denominator of the logarithm in~\eqref{P-Dirichlet} was shown to be equal to~\eqref{he-logw}. Thus, using~\eqref{A1-int} we recast the probability of the layer~\eqref{P-z-def}, which consists of the particles issued from the source at $A$, in the form:
\begin{equation}\label{P-w'}
	P(\mathbf{k})=|w'(A)|^{K}\exp\left\{
	\frac{1}{\hbar^2 K}\left(\int_{l} \int_{l} \log\left|z-z'\right|d^2zd^2z'-\pi\int_{l}\mathcal A(z)d^2z\right)\right\}.
\end{equation}
Interestingly, the growth probability~\eqref{P-w'} can be also obtained within the {\it Dyson gas approach} by considering  the electronic droplet $D^+$ in the non-uniform magnetic field. The droplet grows as the magnetic field varies, and its dynamics is governed by the LG equation~\cite{NormM-growth-1}. In particular, the semi-classical probability to add a single electron to the point $z_0\in \Gamma$ is given by the square amplitude of the one-particle wave function~\cite{QH}:
\begin{equation}\label{psi2}
	|\psi(z_0)|^2=\frac{|w'(z_0)|}{\sqrt{2\pi^3\hbar}}\exp\left\{- \frac{2}{\hbar}\mathcal A(z_0)\right\}.
\end{equation}
By using $K\log|w'(A)|=\int_{l}\log|w'(z)| d^2z$, and assuming that the layer, $l$, is a bump of a single electron, localized at $z_0\in \Gamma$, the double integral in~\eqref{P-w'} disappear~\footnote{In fact, it becomes a cut-off dependent constant.}, and the probability takes the form of~\eqref{psi2}. Thus, $P({\bf k})$ generalizes the result of~\cite{QH} to the case, when the large number of electrons stick to the boundary of the droplet per unit time.

\subsection{Scenario-independent part of probability}

If the growth continue until time $T$, the probability of the growth scenario is given by the product of the conditional probabilities~\eqref{P-w'} for all layers, $l_i=l(i \delta t)$, constituting the final domain $D=\cup_{i=1}^{T/\delta t}l_i$,
\begin{equation}\label{P-dint}
	\mathcal P(\Bbbk)=\mathcal N \mathcal M_D\prod_{i=1}^{T/\delta t}\exp\left\{
	\frac{1}{\hbar^2 K_i}\left(\int_{l_i} \int_{l_i} \log\left|z-z'\right|d^2zd^2z'-\pi\int_{l_i}\mathcal A_i(z)d^2z\right)\right\},
\end{equation}
where the factor,
\begin{equation}\label{Measure}
	\mathcal M_D=\prod_{i=1}^{T/\delta t}\exp\left\{\int_{l_i} \log|w_{i-1}'(z)|d^2z\right\},
\end{equation}
depends on the initial and the final domain only, but not on a particular way to arrive to it. By $w_i(z)$ we denote the conformal map from the exterior of $D_i$ to the complement to the unit disk. To prove the scenario independence of $\mathcal M_D$ we use $\log|w'(A)|=\oint\mu(\zeta,A)\log|w'(\zeta)|$ and  $\p_n G(\zeta,A)=- n\cdot \nabla(A)S(\zeta)$, where $n$ is a unit normal, to recast the exponent in~\eqref{Measure} in the form:
\begin{equation}
	K\log|w'_{i-1}(A)|= \frac{1}{\hbar}\Re\left(\oint_{\Gamma_{i}}S_{i}(\zeta)-\oint_{\Gamma_{i-1}}S_{i-1}(\zeta)\right)\log (w'_{i-1}(\zeta))\frac{d\zeta}{2\sqrt{-1}},
\end{equation}
By changing the summation label, $i-1\to i$, in the latter integral, and using~\eqref{sum-int} to rewrite the discrete summation over time steps in terms of the integral over continuous time, we transform $\mathcal M_D$ as follows:
\begin{equation}\label{log-MD-1}
	\hbar\log \mathcal M_D= \int_0^T\frac{dt}{\delta t}\Re\oint_{\Gamma(t)}S_{t}(\zeta)\log\left(\frac{w_{t-\delta t}'(\zeta)}{w'_{t}(\zeta)}\right)\frac{d\zeta}{2i}+\oint_{\Gamma(T)} S_T(\zeta) \log|w'_T(\zeta)|\frac{d\zeta}{2i},
\end{equation}
where we took into account that the initial domain is a unit disk. Replacing $w'(z)$ by $1/f'(w)$, expanding the ratio of two conformal maps in $\delta t$: $f_{t-\delta t}=f_{t}-\dot f_{t}\delta t$, using $d\zeta=f'(e^{i\phi})d\phi$, and integrating by parts, the first term in the r.h.s.\ of~\eqref{log-MD-1} reads:
\begin{equation}\label{log-MD-2}
	\frac{1}{2\hbar}\int_0^T dt \Im\int_0^{2\pi} \overline {f_t(e^{i\phi})} \dot f_t'(e^{i\phi}) d\phi= \frac{Area}{\hbar} 
	-\frac{1}{2\hbar}\int_0^T dt \int_0^{2\pi}\Im \left(\p_t \overline {f_t(e^{i\phi})} f'_t(e^{i\phi}) \right)d\phi,
\end{equation}
where $Area$ is an area of the annulus $D_T\setminus D_0$. Using LG equation~\eqref{LG-eq} one can prove that the integral in the r.h.s.\ of~\eqref{log-MD-2} equals $Area$. Thus, we arrive to the scenario independent expression:
\begin{equation}\label{A2-fin}
	\mathcal M_D=\exp\left\{\frac{Area}{2\hbar}+\frac{1}{\hbar }\oint_{\Gamma(T)}S_T(\zeta)\log|w'_T(\zeta)|\frac{d\zeta}{2i}\right\}.
\end{equation}

	\subsection{Growth probability, tau-function and K\"ahler geometry}
The double integral over the layer in~\eqref{P-w'}, allows us to relate the growth probabilities with the tau-function of the boundary curves~\eqref{log-tau-def}. First, let us rewrite the double integral,
\begin{equation}
	f\equiv-\frac{1}{\pi^2}\int_{l}\int_{l}\log\left|\frac{1}{z}-\frac{1}{z'}\right|d^2zd^2z'=\frac{1}{2}\delta t_0 \delta v_0+\Re\sum_{k>0}\delta t_k \delta v_k -\frac{1}{\pi}\int_l \mathcal A(z)d^2z,
\end{equation}
in terms of the harmonic moments of the layer, $l(t)=D(t)\setminus D(t-\delta t)$,
\begin{equation}\label{tv-l-def}
	\begin{gathered}
	\delta t_k =-\frac{1}{\pi k}\int_{l}z^{-k}d^2z, \quad \delta t_0 =\frac{1}{\pi}\int_{l} d^2z, \\
	\delta v_k=\frac{1}{\pi}\int_{l} z^k d^2z,\quad \delta v_0 =\frac{2}{\pi}\int_{l}\log|z|d^2z,
	\end{gathered}
\end{equation}
Comparing with~\eqref{tau-series} we conclude,
\begin{equation}
	\frac{1}{2}(\delta t_0)^2 \nabla(A)\nabla(A)\log \tau = f+\frac{1}{\pi}\int_{l}\mathcal A(z)d^2z=\frac12 \delta t_0\delta v_0 +\Re\sum_{k>0} \delta t_k \delta v_k,
\end{equation}
Therefore, the probability~\eqref{P-w'}, can be written in terms of the harmonic moments of the layer~\eqref{tv-l-def}:
\begin{equation}\label{P-l-tv}
	P(\mathbf{k})= |w'(A)|^{K}\exp\left\{
	\frac{\pi^2}{\hbar^2 K}\left(\frac12 \delta t_0 \delta v_0-\Re\sum_{k>0}\delta t_k \delta v_k\right)\right\}.
\end{equation}
Even more compact expression for the probability can be obtained by virtue of the inverse of the conformal map $w(z)$, which is related with the tau-function~\eqref{w-z-def}. Then, the following identity holds~\cite{Tau-calc}: 
\begin{equation}\label{log1-a}
	-\log\left(1-|a|^2\right)=\sum_{k,l>0}\frac{A^{-k}\bar A_m^{-l}}{kl}\frac{\p^2\log \tau}{\p t_k\p\bar t_l},
\end{equation}
where $A=f(1/\bar a)$. Using~\eqref{log1-a} the probability~\eqref{Pw-math} takes the form:
\begin{equation}\label{P-km}
	P({\bf k})= \exp \left\{ - \frac{1}{K\hbar^2}\sum_{k,l>0}\frac{\p^2 \log \tau}{\p t_k\p\bar t_l} \delta t_k \delta\bar t_l\right\},
\end{equation}
where $\delta t_k=K\hbar\, A^{-k}/k$ is a variation of the harmonic moments~\eqref{dt-def} of the domain due to newly attached layer of particles issued from the source at $A$. The quadratic form in the exponent of~\eqref{P-km} determines the metrics on the infinite dimensional complex manifold, such that $t_k$'s are the local coordinates. The metric on this space is K\"ahler, and $\log \tau$ is the K\"ahler potential~\cite{Leon}:
\begin{equation}\label{g}
	g^{k\bar l}=\frac{\p^2\log \tau}{\p t_k \p \bar t_l}.
\end{equation}
The growth of the domain (as the tau-function changes with time), generates a certain geometric flow on the complex manifold. It is of importance to understand the geometric properties of this manifold, and the role of geodesics from the Laplacian growth point of view. This analysis will be done elsewhere.


\section{Hamiltonian structure of the boundary dynamics}\label{4}

The stochastic Laplacian growth, we consider, allows one to determine transitional probabilities between planar domains. The transitions between domains (the {\it growth process}) are described by the stochastic LG equation~\eqref{LGeq-M}, obtained from the variational principle. Remarkably, it turns out possible to reformulate this result in a language of Hamiltonian mechanics, and even to determine the Hamiltonian explicitly.

The time evolution of the Hamiltonian system is defined by the equations:
\begin{equation}\label{H-equations}
	\frac{d {\bm p}}{dt}=-\frac{\p \mathcal H}{\p {\bm q}},\qquad \frac{d {\bm q}}{dt}=\frac{\p \mathcal H}{\p {\bm p}},
\end{equation}
where ${\bm p}$ and ${\bm q}$  are the canonical coordinates, and $\mathcal H=\mathcal H({\bm p},{\bm q},t)$ is the Hamiltonian, which often corresponds to the total energy of the system. As discussed, the logarithm of the transitional probability~\eqref{Pw-math} has clear electrostatic interpretation as an interaction energy between charges, distributed along the unit circle~\eqref{Pmumu}. Considering  this energy as an increment of the total energy of the system, we introduce the Hamiltonian~\footnote{In what follows we set $M=1$ for notation simplicity. However, the construction of the Hamiltonian system can be performed for general $M$ as well}:
\begin{equation}\label{deltaH}
	d \mathcal H(t)=-q(t)\log\left(1-|a(t)|^2\right) dt.
\end{equation}

It seems reasonable to identify the generalized coordinates, ${\bm q}(\zeta,t)$,  with the lengths of \textit{pathlines}, i.e.\ trajectories that the boundary's segment follows during the growth. Clearly, the configuration space is an infinite-dimensional function space labelled by $\zeta\in \Gamma(t)$. It is naturally to define the generalized velocities, $\dot {\bm q}(\zeta,t)$, as the normal velocities of the boundary segments. Thus:
\begin{equation}\label{q-def}
	{\bm q}(\zeta,t)=\int_0^t V_n(\zeta,t')dt',\qquad \dot {\bm q}(\zeta,t)=V_n(\zeta,t).
\end{equation}
In order to determine the canonical momenta, we consider another representation for the Hamiltonian, which can be obtained from~\eqref{P-Dirichlet},
\begin{equation}\label{deltaH-qp}
	d \mathcal H(t)=q(t) \oint_{\Gamma(t)} \mu(\zeta,A)\log\left|\frac{\p_n G_{D(t)}(\zeta,A)}{\p_nG_{D(t)}(\zeta,\infty)}\right|dt,
\end{equation}
and relate it with~\eqref{deltaH} by using the Hamilton's equations. Using~\eqref{A1-def} the numerator of the logarithm in~\eqref{deltaH-qp} can be written in the form:
\begin{equation}\label{fH-1}
	\oint_\Gamma \mu(\zeta,A)\log|\p_n G(\zeta,A)|=\frac{\pi}{2q dt}\oint_\Gamma \mu(\zeta,A)\delta\Phi^-(\zeta),
\end{equation}
where $\delta \Phi^-(\zeta)$ is the variation of the Newtonian potential~\eqref{Phi-def} in $D$ during the growth step due to the source $q$ as the point $A$. As for the denominator, it equals

\begin{equation}\label{fH-2}
	\oint_\Gamma \mu(\zeta,A)\log|\p_n G(\zeta,\infty)|=\oint_\Gamma \mu(\zeta,A) G^-(\zeta,A) 
	+\log|w(A)|+\log\left(1-|a|^2\right),
\end{equation}
where $A=f(1/\bar a)$. By virtue of~\eqref{fH-1} and~\eqref{fH-2} we recast the increment of the Hamiltonian~\eqref{deltaH-qp} in the form:
\begin{equation}\label{dH}
	d \mathcal H=-q\log\left(1-|a|^2\right)d t-\oint_{\Gamma}|d\zeta| \left(G^-(\zeta,A)+\log|w(A)|\right)d {\bm q}
	+\oint_{\Gamma}|d\zeta|\,\dot {\bm q}\, d\left(\frac{\pi}{2q}\Phi^-(\zeta)\right),
\end{equation}
where we took into account $\dot {\bm q}(\zeta,t)|d\zeta|=q\mu(\zeta,A)$ and $d{\bm q}=\dot{ \bm q} dt$. Using~\eqref{H-equations}, and comparing the differential of the Hamiltonian, $d\mathcal H({\bm p},{\bm q},t)=\p_t \mathcal H d t+\p_{\bm p} \mathcal H d {\bm p}+\p_{\bm q}\mathcal H d {\bm q}$, with~\eqref{dH} we conclude that the canonical momentum can be identified with the Newtonian potential of the domain:
\begin{equation}\label{moment-def}
	{\bm p}(\zeta,t)=\frac{\pi}{2q}\Phi^-(\zeta).
\end{equation}
Besides, we also obtain the following set of Hamilton's equations:
\begin{equation}\label{H-eq}
	\begin{gathered}
	\frac{\p\mathcal H}{\p t}=-q\log\left(1-|a|^2\right),\quad \frac{\p\mathcal H}{\p {\bm p}}=\dot {\bm q}(\zeta,t), \\
	\frac{\p\mathcal H}{\p {\bm q}}=-\left(G^-(\zeta,z)+\log|w(A)|\right),
	\end{gathered}
\end{equation}
where $G^-(\zeta,z)+\log|w(A)|=\dot {\bm p}(\zeta,t)$ as follows from the Laplacian growth equation~\eqref{Phi-G}. Thus, the boundary's dynamics is Hamiltonian and the time dependent Hamiltonian, $\mathcal H(t)$, can be written in the form:
\begin{equation}\label{H-def}
	\mathcal H(t)=-\int_0^t dt'\sum_{m=1}^{M}q_m(t')\log\left(1-|a_m(t')|^2\right).
\end{equation}

The following rough analogy with quantum mechanics might be helpful. The quantum systems stays in a single state indefinitely long, if the Hamiltonian does not depend on time. Transitions between states occur, when the time-dependent perturbation is applied to the system. The same reasoning applies to the stochastic Laplacian growth, if one identifies the Hilbert space of the theory with the equivalence classes of domains, such that the clusters which differ by area only are equivalent. In other words, the Hilbert space of stochastic LG is a projective space. The growth, governed by a single hydrodynamical source at infinity, keeps the system in the same state, as the harmonic moments, $t_k$ with $k>0$, are kept fixed. It should  be the {\it time-independent Hamiltonian}, $\mathcal H_0$, associated with this process. Transitions between the states in the Hilbert space occur, when the time-dependent perturbation changes the harmonic moments of the cluster. The associated {\it time-dependent Hamiltonian}, commonly denoted by $\mathcal H_{int}(t)$ in quantum mechanics, is~\eqref{H-def}.


	\section{Possible connection to the Liouville field theory}\label{5}


As a final matter we address a possible relation between the probabilities for occurring fluctuations in the interface dynamics in LG, and the semi-classical limit of the correlation functions in the Liouville conformal field theory on a pseudosphere. Clearly, the Robin's function~\cite{Robin}, $G^-(w,w)=-\log|1-w\bar w|$, for the Dirichlet problem inside the unit disk in the $w$-plane, satisfies the classical Liouville equation:
\begin{equation}\label{Leq}
	\p\bar \p G^-(w,\bar w)=e^{2 G^-(w,\bar w)}
\end{equation}

In differential geometry the Liouville equation appears as a partial differential equation for the conformal factor, $\exp \varphi(w,\bar w)$, of a metric $ds^2=\exp \varphi(w,\bar w)|dw|^2$ on a two-dimensional surface of constant Gaussian curvature. In particular, if  $\varphi(w,\bar w)=2G^-(w,\bar w)$, the Liouville equation~\eqref{Leq} describes the geometry of the so-called Lobachevskiy plane, or pseudosphere, which is a two-dimensional surface with a constant negative curvature $R_{ds^2}=-4$. It can be realized as the Poincar\'e disk model, such that the points of the geometry are inside the unit disk $B=\{w\in\mathbb C:|w|<1\}$.
The metrics $ds^2=\exp\varphi(w,\bar w)|dw|^2$ is given by the solution to the Liouville equation~\eqref{Leq}: $\varphi(w,\bar w)=-\log(1-w\bar w)^2$.
Since the Liouville field, $\varphi(w,\bar w)$ appears as the conformal factor of a metric tensor, under holomorphic coordinate transformations, $w\to z=f(w)$, it transforms as $\varphi\to \tilde\varphi(w,\bar w)=\varphi(z,\bar z)+2\log|f'(w)|$. This transformation law relates the growth probabilities,~\eqref{P-z-def} and~\eqref{Pw-math}, directly. In terms of the Liouville field, the probability of the layer in the $w$-plane~\eqref{Pw-math} appears as a product of conformal factors:
\begin{equation}\label{P-Liouville}
	P(\mathbf{k})=\mathcal N\prod_{m=1}^Me^{-(K_m/2)\varphi(a_m,\bar a_m)}.
\end{equation}

This expression can be treated as the semi-classical limit of a certain correlation function in the Liouville field theory~\cite{ZZ-1}. 
Equation~\eqref{Leq} appears as the Lagrange-Euler equation for the Liouville action:
\begin{equation}\label{S_L}
	S_L[\varphi]=\frac{1}{8\pi b^2}\int_{|w|<1} \left(2|\p\varphi|^2+8e^{\varphi}\right)d^2w,
\end{equation}
where the integral is taken over the unit disk. The coefficient $b$ in~\eqref{S_L} is an important parameter of the theory. 
The Liouville action determines the correlation functions of the exponential fields, $\exp \alpha\varphi(z,\bar z)$,
where the parameters $\alpha$ label conformal classes of exponential operators~\cite{ZZ-2}. In the semiclassical limit, $b\to0$, the saddle-point approximation can be used. There are two classes of primary operators: 1) the ``heavy'' operators, $\alpha_m\sim b^{-1}$, non-trivially effect on the saddle point, and 2) the ``light'' operators, $\alpha=\mu b$ (where $\mu$ is a constant), influence neither the classical solution nor the one-loop correction. In the pseudospherical geometry, the semiclassical behaviour of the correlation function of ``light'' fields is
\begin{equation}\label{Cf-cl}
	\left\langle \prod_{m=1}^M e^{\mu_m\varphi(a_m,\bar a_m)}\right\rangle\approx e^{-S_L[\varphi_{cl}]}\prod_{m=1}^M e^{\mu_m\varphi_{cl}(a_m,\bar a_m)}.
\end{equation}
We notice a similarity between the growth probabilities~\eqref{P-Liouville} and the correlation functions of the ``light'' fields~\eqref{Cf-cl} in the semiclassical limit.
One can argue, that the Liouville field theory appears quite naturally in the context of the stochastic Laplacian growth.
However, only computation of semiclassical corrections to $P({\bf k })$ might provide a deep insight in the possible link between Laplacian growth and conformal field theories first noted in~\cite{LargeN}.

	\section{Conclusions}

In this paper we considered the deterministic (classical) limit of the discrete stochastic Laplacian growth, which reveals a strong connection between LG and diffusion limited aggregation. We related the stochastic growth with generation of virtual sources, which are the ``fingerprints'' of the domains.
The Hastings-Levitov stochastic dynamics~\cite{Levitov} is a one among many others stochastic growth scenarios, where only the bumps (not layers) stuck to the interface per time unit. Simple probabilistic arguments allow us to introduce the probability of the generated patterns, so providing an analytic framework to study stochastic pattern formation.

The growth probability of a single layer has a form of Kullback-Leibler entropy, and can be rewritten as a certain functional on the layers, which has remarkable relations with the tau-function for the boundary curve. It allows us to connect the stochastic Laplacian growth with the random matrix theory. For the classical LG this relation has been first noticed in~\cite{Tau-calc}, and used to compute density-density correlation functions in~\cite{ZabCor}.
In the subsequent publications we will address the matrix model interpretation of the growth probabilities in detail.

Remarkably, we managed to relate the layers probabilities~\eqref{P-km} with the geometric characteristics (metric tensor~\eqref{g}) of the infinite-dimensional K\"ahler manifold, spanned by the external harmonic moments of the growing domain. The evolution of the domain generates certain geometric flow on the manifold, as the tau-function changes with time.

We proved the Hamiltonian structure of the interface dynamics, so that the famous Laplacian growth equation (more precisely \eqref{Phi-G}) turns out to be the Hamilton's equation for a certain dynamical system. We also determined the time-dependent Hamitonian, which generates transitions between different states (equivalence classes of contours) in the Hilbert space of the theory.

Finally, the observed relation between growth probabilities of layer and certain correlation functions in the Liouville field theory in a pseudosphere  awaits further clarification. A very few remarks on the related subject can be found in~\cite{LargeN}, where the next order corrections to the free energy of the Dyson gas in the large $N$ limit were related with the spectral determinant of the Laplace-Beltrami operator for the unit disk. This observation suggests interesting links with the conformal field theory.

\end{document}